\documentclass{article}
\usepackage[utf8]{inputenc}
\usepackage{graphicx}
\usepackage{hyperref}

\title{A portable muon telescope based on small and gas-tight Resistive Plate Chambers}
\author{Sophie Wuyckens, Andrea Giammanco, Pavel Demin, \\  Eduardo Cortina Gil\\
\\
Centre for Cosmology, Particle Physics and Phenomenology,\\ Universit\'e catholique de Louvain, \\ Louvain-la-Neuve, B-1348, Belgium}

\begin{document}

\maketitle

{\raggedleft CP3-18-36 \\
}

\begin{abstract}
We report on the first steps in the development of a small-size muon telescope based on glass Resistive Plate Chambers with small active area ($16\times 16~{\rm cm}^2$). 
The long-term goal of this project is to focus on applications of muography where the telescope may have to be operated underground and/or inside small rooms, and in challenging logistic situations. 
Driving principles in our design are therefore compact size, light weight, gas tightness, and robustness. 
The first data-taking experiences have been encouraging, and we elaborate on the lessons learnt and future directions for development.
\\
\\
{\it Proceedings of the ``Cosmic-Ray Muography'' meeting of the Royal Society, 14-15 May 2018 at the Kavli Royal Society Centre, Chicheley Hall, Newport Pagnell (UK). Submitted to Philosophical Transactions of the Royal Society A.}
\end{abstract}

\section{Introduction}
Muography is an imaging technique that exploits the abundant natural flux of muons from cosmic-ray interactions in the atmosphere~\cite{procureur-review}. It was first applied for practical purposes in the 1950s to measure the rock overburden above a tunnel~\cite{george}, and in the 1960s in a search for hidden chambers in the Pyramid of Chephren in Giza~\cite{alvarez}. 
In the last decade, several teams have developed large-area muon telescopes for applications in volcanology (for a review, see for example Ref.~\cite{review-saracino-carloganu}). 
Muography also finds applications in the imaging of smaller targets, such as in nuclear waste or spent fuel monitoring~\cite{jonkmans,firenze-glasgow,perry,raffaello-waste} and archaeology~\cite{scanpyramid,echia,tharros}.
This article reports on the development of a first prototype of a portable telescope based on small and gas-tight glass Resistive Plate Chambers (gRPC)~\cite{grpc}, aimed at small targets in contexts of challenging logistics.

\section{Experimental set-up}
\label{sec:setup}

This project is based on the same gRPC design as in the CMS experiment at the Large Hadron Collider~\cite{cms-rpc} and the Hadronic Digital Calorimeter developed within the CALICE collaboration for usage at future accelerators~\cite{calice}. The same design is used in muography by the TOMUVOL collaboration for applications in volcanology~\cite{tomuvol}. 
While large-area gRPCs, of O($1~{\rm m}^2$), are needed in these High-Energy Physics and volcanology use cases, we use an alternative design with smaller active area ($16\times 16~{\rm cm}^2$). 
Their compact size, low cost, excellent space and time resolutions, and the stability of their response make gRPCs an attractive choice for muography. 

A specific driving principle for our design is gas tightness, to make the apparatus more portable and solve the safety and logistic issues that usually affect gas detectors operated underground and/or inside small rooms. We also aim at ease of transportation of the entire set-up, including its electronics.

\begin{figure}[!h]
\centering\includegraphics[width=4.5in]{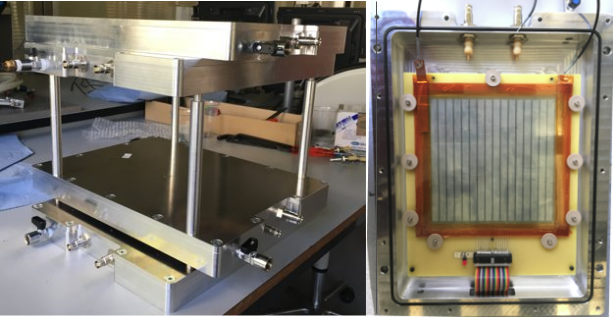}
\caption{Left: Our muon telescope. Right: One of the mini-gRPC detectors inside its casing.}
\label{fig:setup}
\end{figure}

The set-up documented in this article consists of four planes of gRPC detectors, arranged as in Fig.~\ref{fig:setup} (left). We use removable vertical bars in order to create a distance of 14.8~cm between the second and third plane. 
Each detector, as shown in Fig.~\ref{fig:setup} (right), has 16 sensitive strips, with 1~cm pitch and width 0.9~cm, and is hosted in an aluminum box with dimensions $38\times29\times 3.7~{\rm cm}^3$, designed to be \textbf{gas tight}. 
A single detector plane weighs 6.5 kg. 
The first and third detector are oriented orthogonally to the second and fourth, in order to provide bidimensional information ($x$ and $y$ orientation) at the top and at the bottom of the telescope.

We operate the detectors with a mixture of freon (95.2\%), SF$_6$ (0.3\%) and isobuthane (4.5\%), with a pressure slightly higher than the atmospheric one.
Before filling the detector casings with the operating gas, tightness tests were performed by measuring the leakage rate after creating vacuum and looking for leaking points with helium.  
The initial tightness tests spotted several leakage points, but after a few fixes we were able to achieve a leak rate of $10^{-9}~{\rm mbar\cdot l/s}$. 

The glass sheets used as electrodes in our detectors were cut in a square shape of $20\times 20~{\rm cm}^2$, 1.1 mm thick. As a resistive coating, we painted the glass with a colloidal dispersion of antimony-doped tin oxide in water (20\% powder, 80\% water) mixed with an equal quantity of methanol. 
To ensure a uniform gas gap, nine round edge spacers made in PEEK (polyether ether ketone) are used between the glass sheets. 
We use front-end electronic boards (FEB) identical to those used for the RPCs of the CMS experiment~\cite{bruno,fee}. 
Each FEB has 4 front-end modules containing 8 channels each; therefore we use one FEB for two detectors. 
Each channel consists of an amplifier with a charge sensitivity of $2~{\rm mV/fC}$, a discriminator, a monostable and a LVDS driver. 
Once a pulse is detected, a time window of 80~ns opens and all pulses that happen during this period of time are added (logic OR of all channels).

\section{Data collection, data analysis, and results}

Two data collection campaigns are described here. 
The first one, with an incomplete set-up, was performed outdoors, with a challenging logistic situation that we consider representative of potential use cases of this telescope for muography applications. 
The second test was performed in the comfortable environment of our laboratory, after having addressed the issues identified during the first test. 
This order is admittedly unusual (it is costumary and logical to perform a first complete test in controlled conditions before sending a detector in the wild), but it was motivated by the coincidental opportunity for one of the authors of this study to take part to "UCL to Mars 2018"~\cite{ucl2mars}, a two-week long life-on-Mars simulation at the Mars Desert Research Station (MDRS) in the Utah desert (USA) in March 2018. Participants were requested to design a scientific experiment to be performed at the MDRS, and this opportunity was seized to verify the portability and robustness of our prototype detectors.

\subsection{A first test with real-life logistics}

As mentioned above, the first campaign was led in March 2018 in the MDRS. 
Only two out of the four detectors were ready at that time and were shipped to the USA. 
Nominally, the gas mixture differed from Sec.~\ref{sec:setup} only in the presence of argon instead of freon. 

Two crucial issues affected the collection of these data:  
the fractions of SF$_6$ and isobuthane were not as expected, resulting in numerous sparks and the impossibility of a satisfactory optimisation of high voltage and thresholds, 
and there was a very large ambient noise, picked up from a power generator. 
But although the data collected during this campaign were useless for actual studies, this experience was not in vain, as we could verify the main intended features of this detector design:

\textbf{Gas-tightness}: We verified that the pressure in the detector casings did not decrease significantly on a time scale of a month.

\textbf{Portability and compactness}: The two detectors have been shipped from Louvain-la-Neuve (Belgium) to a remote location in the Utah Desert (USA) using a wooden box of size $60\times 40\times 69~{\rm cm}^3$ that also contained the electronics rack (high voltage and readout system), a transformer and useful tooling for a total of 37.5~kg. At the MDRS, a single person was able to move and operate the detectors inside and outside of the station thanks to the ease of carriage. 

\textbf{Robustness}: The detectors survived a round trip between Belgium and USA without any problems, and are still operational at the time of writing. Data were collected outdoors even in conditions of strong wind or high temperatures.

\subsection{Optimisation and new data collection}

Upon return of the detectors to Belgium, we addressed the issues that affected the quality of data at the MDRS. 
We realised that the problem with the gas behaviour could be traced back to an issue with the gas mixing system in our laboratory, which will need a refurbishment. 
Therefore, we refilled all four detectors with the gas mixing system of Ghent University, obtaining satisfactory results from some quick performance tests. 
We also realised that the problem with environmental noise was due to grounding issues. 
After that, we continued the studies in our laboratory at UCL.

\begin{figure}[!h]
\centering\includegraphics[width=0.45\textwidth]{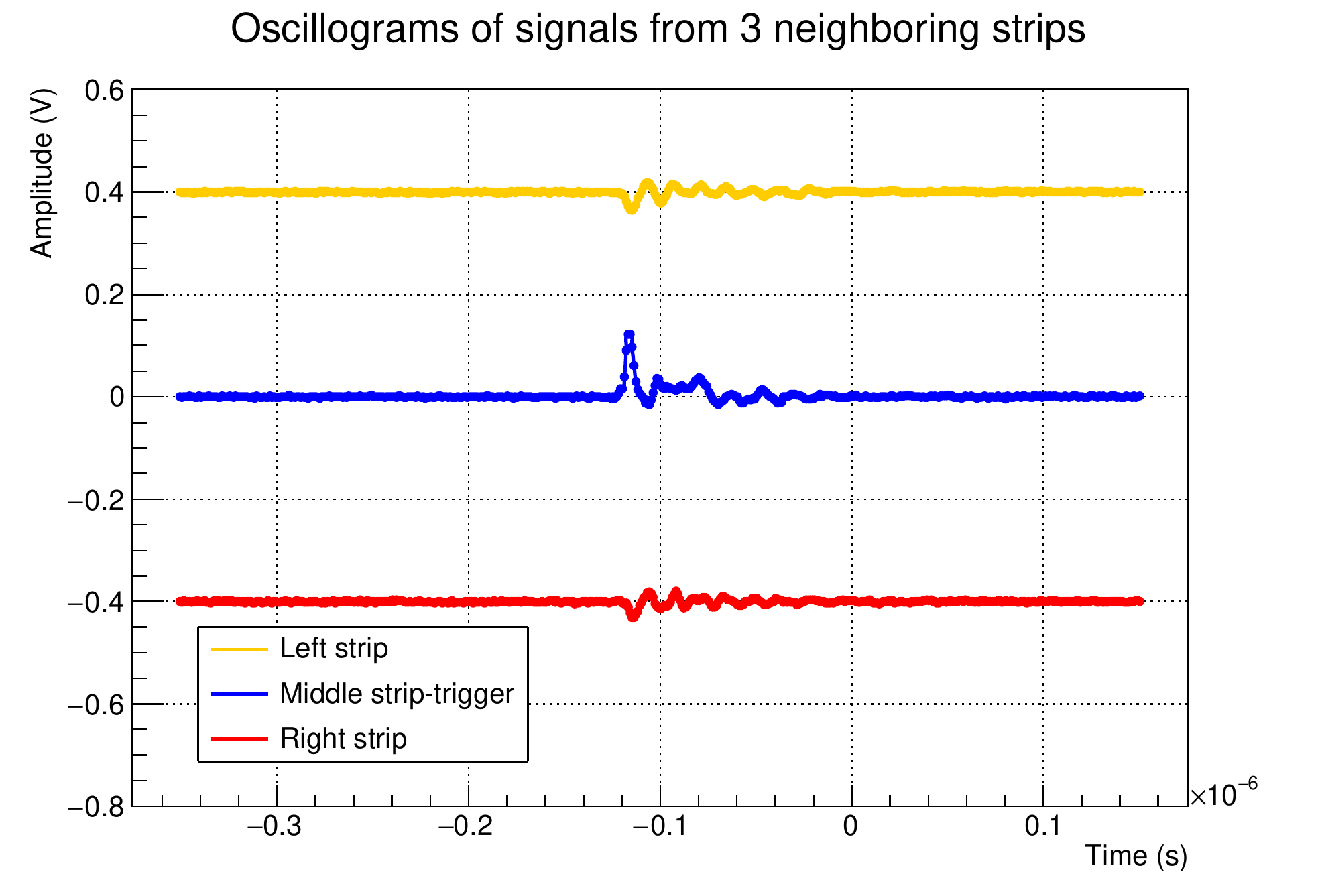}
\centering\includegraphics[width=0.45\textwidth]{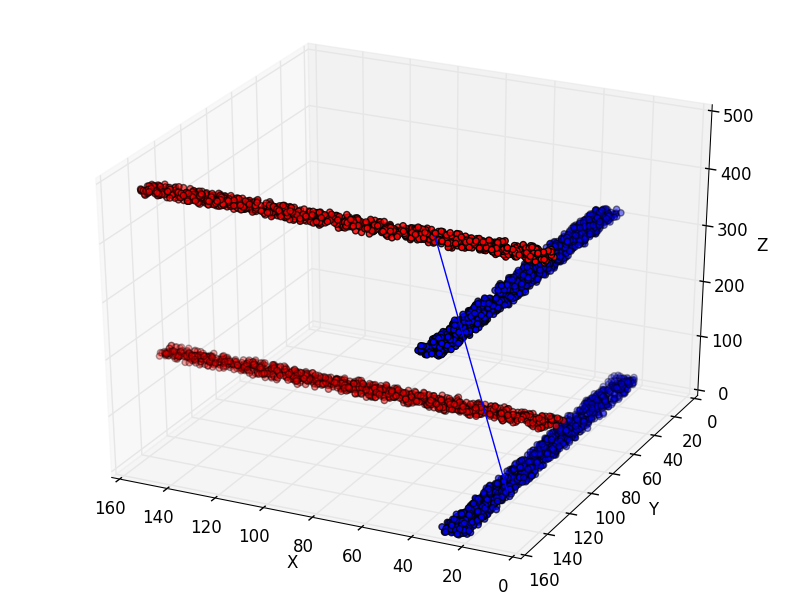}
\caption{Left: Signal induced by the passage of a muon on three neighboring strips, where the trigger is given by the middle one. Right: Reconstructed muon trajectory.}
\label{fig:oscillo}
\end{figure}

Figure~\ref{fig:oscillo} (left) shows the signal induced by the passage of a muon on three neighboring strips. A signal of positive polarity is visible in the strip that gives the trigger, while small negative-polarity signals are induced in the neighboring strips by cross-talk. 
The signal is followed by reflections due to wrong impedance matching in the cables. 
Figure~\ref{fig:oscillo} (right) shows an illustrative event from the data, with the reconstructed trajectory of a muon giving signal in all planes.

\begin{figure}[!h]
\centering\includegraphics[height=1.5in]{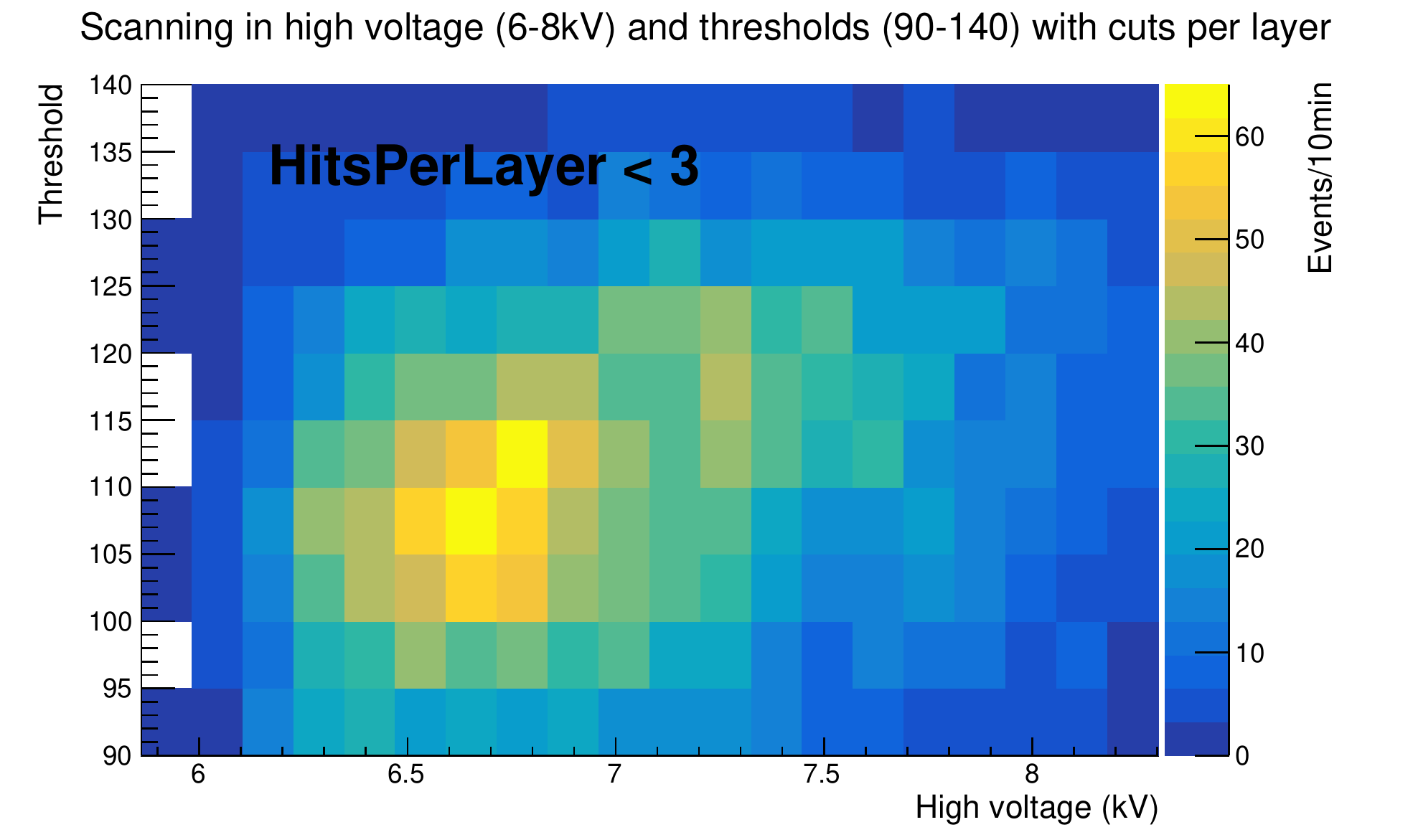}
\centering\includegraphics[height=1.5in]{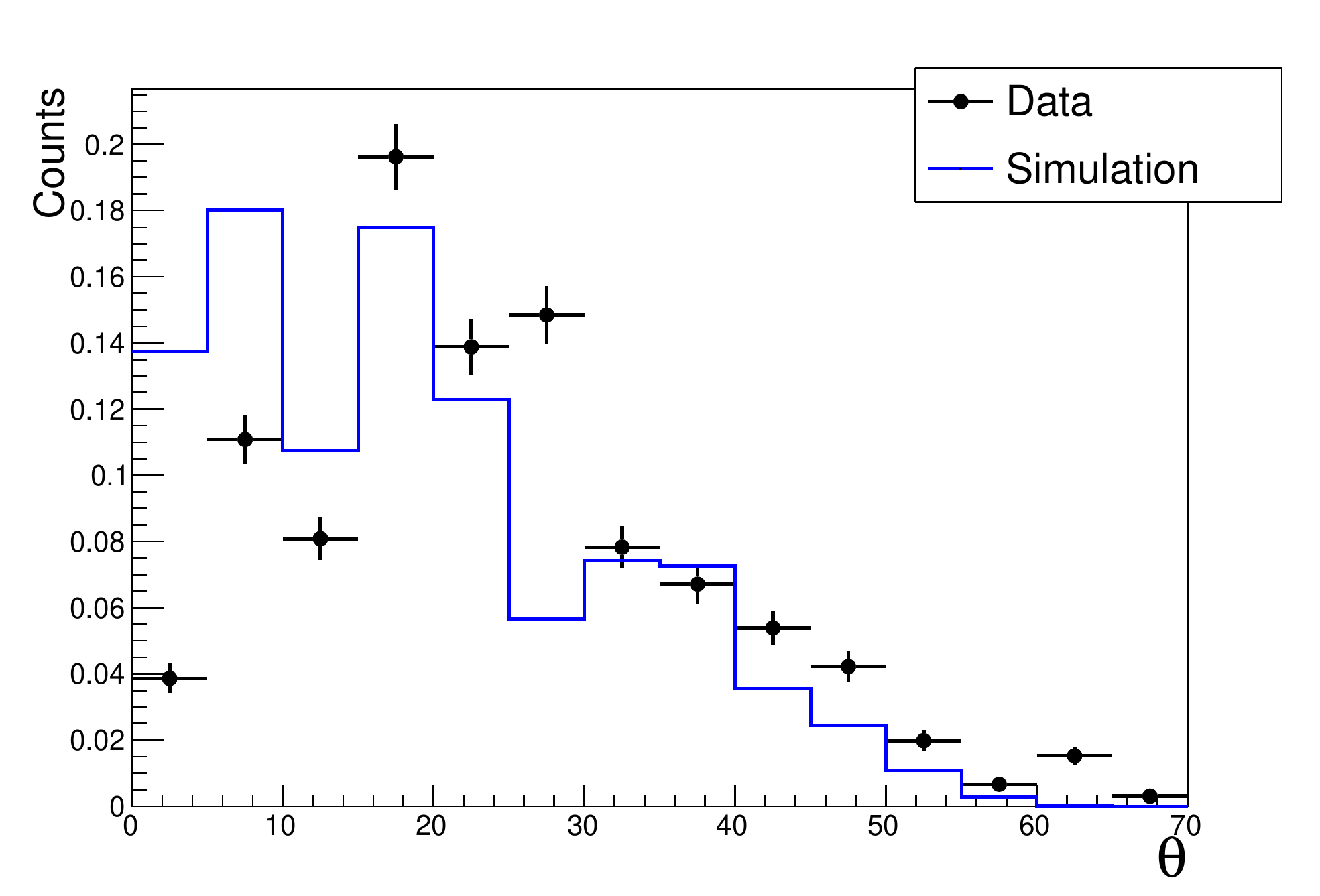}
\caption{Left: Number of events where at least one strip and less than three strips per detector give a signal, as a function of high voltage and threshold. Right: Distribution of the zenith angle reconstructed with our final set-up, collecting data for six days, compared to the result of a simulation. Plots created with ROOT~\cite{root}.}
\label{fig:scan}
\end{figure}

To find the optimal working point in terms of high voltage and thresholds, the following automated procedure was set up.
The high voltage was scanned between 5.9 and 8.3~kV in steps of 12 V. 
The threshold on signal amplitude, although it can in principle be varied indipendently per FEB module, has been set equal for all channels and all detectors, and for each value of the high voltage it was scanned between 90 and 140 (arbitrary units) in steps of 5. 
For each value of high voltage and threshold, data were taken for 10 minutes. 
The complete optimisation procedure took therefore around 33 hours of data collection.
Figure~\ref{fig:scan} (left) shows the event counts as a function of high voltage and threshold, when requiring that at least one strip, but less than three, per detector are above threshold. By taking the maximum from this plot, we conclude that the optimal working point corresponds to a high voltage of 6.6-6.7~kV and a threshold of 105-110 (arbitrary units). 

The first and last strip in each detector are found to be very noisy and are excluded from further analysis. 
Figure~\ref{fig:scan} (right) shows the distribution of the zenith angle reconstructed in events with exactly one hit per detector, excluding the extreme strips, after six days of data collection. This distribution follows the qualitative expectation obtained from a crude parametric simulation of the detector set-up, but discrepancies are observed especially at low values of this angle.

\section{Conclusions, lessons learnt, and outlook}

A first prototype for a new type of telescope for muography applications, based on mini-gRPC detectors, has been developed following the requirements of small cost and portability, where the latter is to be intended not only in terms of dimension and weight but also by ensuring gas tightness and ease of transportation of the full set-up, including electronics.

This first experience has been encouraging, as we were able to design, build from scratch, and operate our muon telescope in a very short time, verifying the robustness and portability of the set-up and achieving reasonable performances. 
In the process, we learned several lessons. 
We need to invest on a reliable gas mixing system, and we need to measure the efficiency of our detectors as a function of time over long time scales; in fact, the quality of the gas can degrade over time, even in absence of significant leakages, due to the out-gassing of the detector components~\cite{procureur-private}.

After addressing the aforementioned issues, we plan to build a new prototype to move further from proof of principle towards real applications. The following developments are considered.

The intrinsic spatial resolution of RPCs can easily be better than O(1~mm), therefore thinner strips would allow us to perform high-resolution muography. These resolution scales have been reached, for similar applications, by other gas detectors (see for example Refs.~\cite{scanpyramid,watto,regard}) while being out of reach for scintillator-based telescopes (see for example Refs.~\cite{echia,tharros}). 
The natural drawback of an increase in the density of channels is a corresponding increase in the complexity and cost of the electronics. 
In order to read-out more channels at a relatively small cost, we consider replacing our CMS FEBs with HARDROC chips~\cite{hardroc}. These are 64-channel front-end ASICs designed to read-out fast ($< 1~{\rm ns}$) and short ($< 10~{\rm ns}$) current pulses, that can be embedded inside the detectors without any external circuitry. 
Gains and thresholds can be tuned per channel, to allow for equalisation of the response. 
In addition, an appealing cost-saving option for muographic applications, where the rate of signals is relatively small, is a clever use of multiplexing in order to read a large number of strips with as few channels as possible. A method that may be appropriate for our low-rate use cases is the so called ``genetic multiplexing''~\cite{multiplexing}, where the strips can be made as thin to ensure that most hits give signal in two neighboring strips and this redundancy can be exploited to compress the information in an unambiguous way. 

In addition to the intrinsic spatial resolution, RPCs are also characterised by an excellent time resolution, O(1~ns); while this is not exploited in our current set-up, achieving this resolution and increasing the distance between the planes would allow, in muography applications, to discriminate the signal muons from the backgrounds coming from a variety of sources.

And finally, we plan to keep our set-up fully modular (as it is now), in order to add more planes and adjust the distances at ease.

\section*{Acknowledgements}
The authors gladly acknowledge the crucial help of Daniel Michotte and Nicolas Szilasi in the electronics and mechanical design aspects of this project, and the usage of resources of the mechanical workshop of the 
Universit\'e catholique de Louvain. Jean-Pierre Clare (Centre de Recherches du Cyclotron, CRC) helped us with the leak tests reported in this article. The original proposal for mini-gRPCs was put forward by Michael Tytgat from Universiteit Gent, and we are indebted to Alexis Fagot and Antoine Pingault from Michael's team, who shared with us their expertise in gRPC development. We borrowed the CMS FEB from them and profited from the gas mixing system of their laboratory at the Department of Physics and Astronomy of Universiteit Gent, where we also performed some tests. SW and AG are grateful for the financial contribution to their participation to the Theo Murphy International Meeting of the Royal Society on ``Cosmic-Ray Muography'' where this work was presented in May 2018, received through the ``seed fund'' of the Centre for Cosmology, Particle Physics and Phenomenology, convention IISN 4.4503.17-F of the Fonds National de la Recherche Scientifique.



\end{document}